\documentclass{arxiv-svjour3}                     
\smartqed  

\usepackage{natbib}
\usepackage[]{epsfig}
\usepackage{graphicx}
\usepackage{url}
\usepackage{tabularx}
\usepackage{booktabs}
\journalname{Scientometrics}
\begin{document}

\title{A Fitness Model for Scholarly Impact Analysis}

\author{Weimao Ke
}

\institute{Weimao Ke \at
              College of Information Science and Technology, Drexel University \\
              Tel.: +1 (215) 895-5912 \\
              Fax: +1 (215) 895-2494\\
              \email{wk@drexel.edu}
}

\date{Received: 2012 / Accepted: 2012}

\maketitle

\begin{abstract}
We propose a model to analyze citation growth and influences of fitness (competitiveness) factors in an evolving citation network. Applying the proposed method to modeling citations to papers and scholars in the InfoVis 2004 data, a benchmark collection about a 31-year history of information visualization, leads to findings consistent with citation distributions in general and observations of the domain in particular. Fitness variables based on prior impacts and the time factor have significant influences on citation outcomes. We find considerably large effect sizes from the fitness modeling, which suggest inevitable bias in citation analysis due to these factors. While raw citation scores offer little insight into the growth of InfoVis, normalization of the scores by influences of time and prior fitness offers a reasonable depiction of the field's development. The analysis demonstrates the proposed model's ability to produce results consistent with observed data and to support meaningful comparison of citation scores over time. 
\end{abstract}

\setcounter{secnumdepth}{-1} 

\section{Introduction}

Citation frequency is a basic indicator of the use and usefulness of scientific publications \citep{pritchard:bib}. Citation analysis has been commonly used to evaluate scholarly productivity and impact \citep{garfield:citation,cronin:citation}. However, due to human subjectivity in citation behaviors and a wide spectrum of factors involved in a scholar's decision to include a reference, citation frequency is not an unambiguous quantity for objective evaluation of scholarly communications \citep{nicol:citation,weightIF}. According to \citet{garfield:citation}, citation scores are associated with many variables beyond scientific merit. 

It has been recognized that citation growth is a process of cumulative advantage, in which ``success seems to breed success'' \citep[p. 292]{price:bib}. Early players are likely to dominate in gaining citations given the advantage of entry time. In terms of \citet{science:rdnnet}, complex systems such as a citation network demonstrate {\it scale-free} properties as a result of network growth with {\it preferential attachment}. Specifically, when a node enters a {\it scale-free} network, it is more likely to connect to (cite) those that have been more strongly connected (highly cited). The scale-free model nicely explains distributions of connectivity that decay with a power law function, commonly observed in real world networks such as the world wide web and citation networks \citep{popular:paper,smallworld,barabasi2009,citeDist}. 

Despite its model simplicity and effectiveness in regenerating related distributions, the {\it preferential attachment} mechanism in the scale-free model only represents partial reality. While many real-world connectivity distributions show a long tail, they are rarely perfect power-laws that are {\it free} of scales. Due to constraints such as aging and limited capacities to receive new connections, certain categories of these networks demonstrate single-scale or broad-scale characteristics \citep{Amaral2000}. Broad-scale structures such as co-authorship networks demonstrate a power-law region followed by an exponential or Gaussian cutoff because of individual capacities to collaborate. 

In addition, {\it preferential attachment} alone does not sufficiently depict the reality given the common observation that competitive latecomers do have chances to break the loop and play important roles in growing network communities. In written communications, an article with great scientific metric may attract lots of citations even if it is published lately \citep{popular:paper}. This recognition of {\it competitiveness}, in addition to the time factor in preferential attachment, has triggered research on new models in network science. 

According to \citet{fitness}, a node's growth in connectivity in a network depends on its {\it fitness} to compete for links. Fitter nodes have the ability to overcome highly connected nodes that are less fit. In the fitness model, entry time as well as factors associated with a node's competitiveness (fitness) account for its ultimate connectivity. A fitness network demonstrates not only the rich-get-richer effect (dominance of early players) but also the fitter-get-richer phenomenon (opportunities for latecomers to surpass the established). 

The ideas of network growth, preferential attachment, and fitness have important implications in citation analysis. We have acknowledged that raw citation count is not a fair vehicle for scholarly impact evaluation. Particularly, time is a factor that likely hinders meaningful comparison of papers published in different years. While one may suggest the use of yearly averages to normalize citation scores for a comparative evaluation, research has clearly indicated that citation growth is not a linear function of time \citep{sumner:refage,gupta:cite-age,science:rdnnet,fitness,aging}. How to isolate the influence of time in citation analysis requires close examination of this relationship. 

Furthermore, we reason that {\it fitness} is a very broad notion and, in the context of citation analysis, potentially represents a variety of constituent variables. To understand quantitatively the process of papers\footnote{We use papers, articles and publications interchangeably. In the data used for this study, a paper may refer to a research article, a book chapter, or a book.} competing for citations requires a mathematical model in which the notion of {\it fitness} is integrated and can be factorized into related variables in citation data. 

In the light of {\it preferential attachment} and {\it fitness}, this research aims to build a simple, general model to quantify citations and analyze scholarly impacts in evolving citation networks. The model will integrate time as well as related fitness factors in the modeling and offers a means to single out contributions (bias) on citation scores for comparative analysis. We will conduct a case study to validate the proposed model and to demonstrate its utility in the evaluation.

\section{Proposed Model}

We present a fitness model to analyze citation distributions over time. The purpose of this modeling is to quantitatively offer insight into citation characteristics and evolving patterns in various domains. While its applicability can be verified with real data, the model will incorporate important variables and take into account their relations in contributing to scholarly impact. By quantifying individual factors' contributions, we can estimate key parameters in the model and obtain important quantitative descriptors about the development of a domain in question. 

We describe the proposed model by introducing three key aspects in the analysis. It is apparent that the number of citations a paper has received reflects several factors in the following respects: 1) quality, merit, and contribution of research presented in the paper; 2) attractiveness of the paper due to existing influences of its authors and publication venue; and 3) age of the paper which allows for the accumulation of citations over time. We model citation-based scholarly impact using these (abstract) factors and elaborate on model formulation below. 

\subsection{Citations over Time}

\begin{figure}[ht]
\centering
\begin{tabular}{cc}

\begin{minipage}{2.2in}
\centering
\epsfig{file=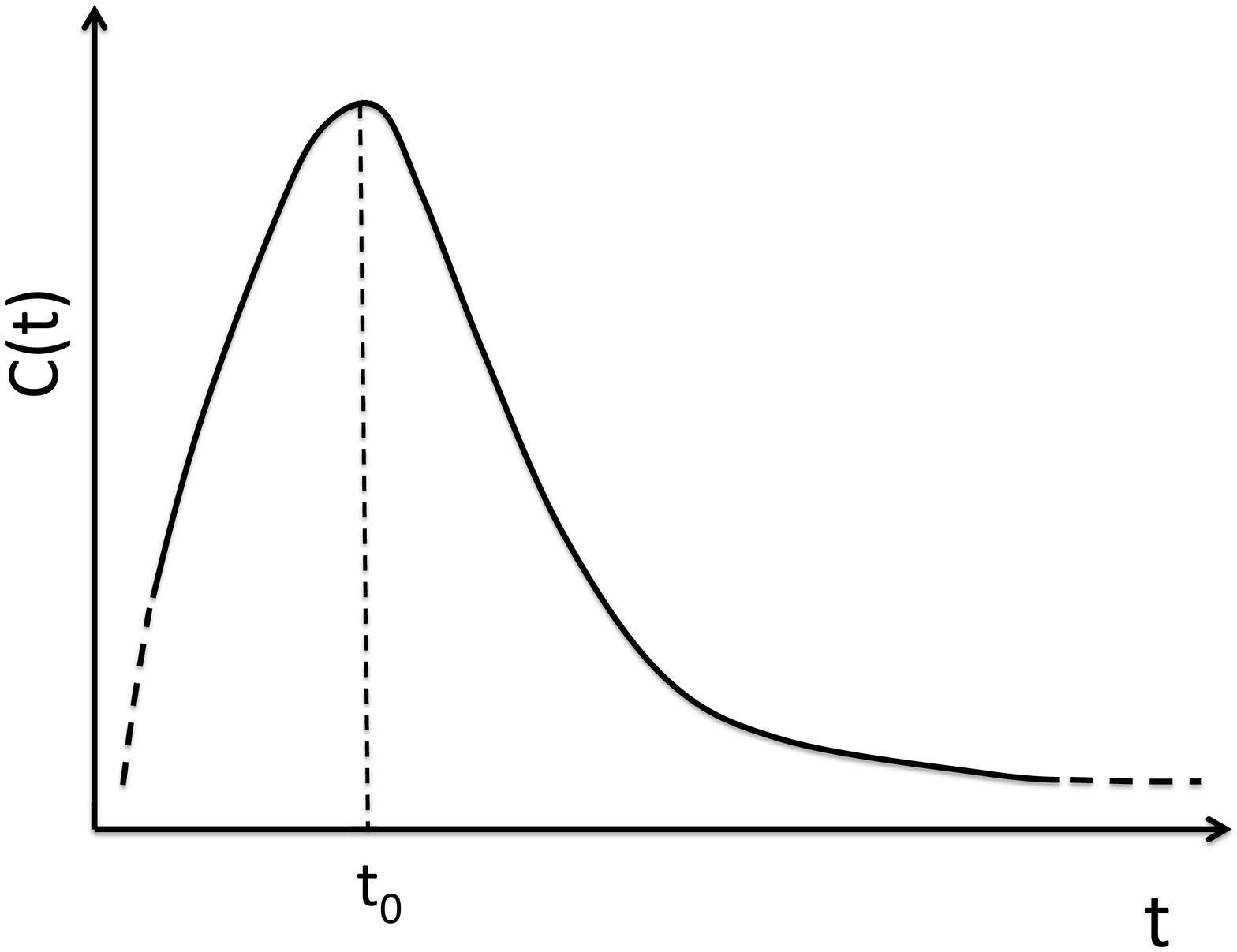,width=2.2in}
\end{minipage}
&
\begin{minipage}{2.2in}
\epsfig{file=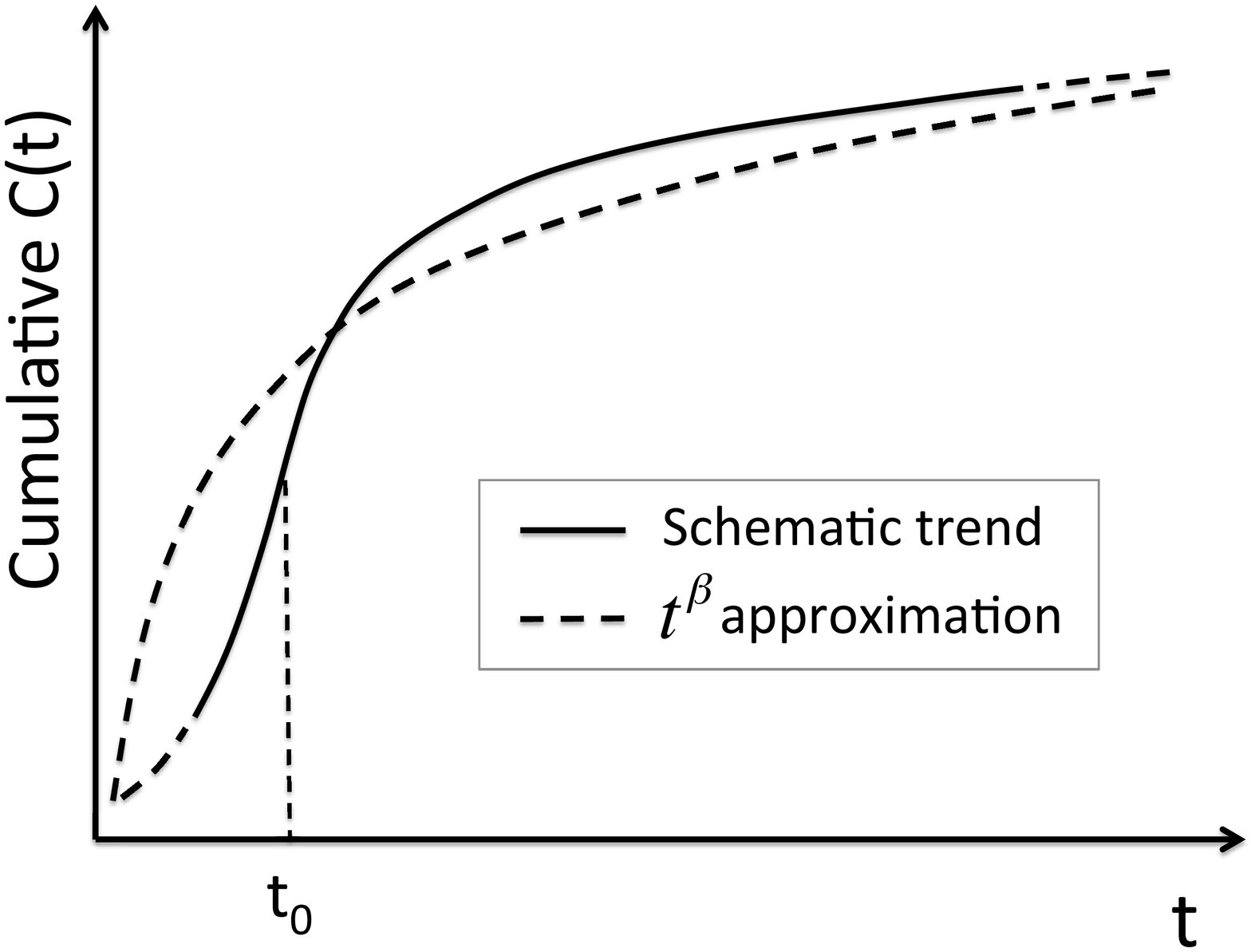,width=2.2in}
\centering
\end{minipage}
\\
(a) Citations &
(b) Cumulative citations
\\
\end{tabular}
\caption{Citations $c(t)$ vs. time $t$. (a) illustrates a schematic citation distribution over time; (b) presents the cumulative distribution and a $cc(t) \propto t^\beta$ approximation of the schematic trend. Compare to \cite{gupta:cite-age}.}
\label{fig:ct}
\end{figure}

Research has identified some common patterns about how citations accumulate over time. According to \citet{gupta:cite-age}, a citation decay curve consists of two parts: an increase of citations during first couple of years followed by gradual decline of citations when the paper gets older, as shown in Figure \ref{fig:ct} (a). The cumulative trend is illustrated in Figure \ref{fig:ct} (b). Similar patterns have been found in related studies such as \citet{sumner:refage,aging}. 

Viewing papers as nodes and citations as directed arcs connecting the nodes, network science research has provided important methods and tools to model citation frequency based on connectivity probabilities in an evolving citation network. The scale-free model, among others, provides insights into the basic mechanisms behind power-law degree distributions commonly observed in a wide spectrum of real networks. It models the outcome of such a distribution based on two simultaneous processes, namely network growth and preferential attachment. The original scale-free model proposed by \cite{science:rdnnet} relies on a probability function linear to existing connectivity. That is, the likelihood that a new node connects to an existing node $i$ (paper $i$) is proportional to node $i$'s degree (the number of citations paper $i$ has already received). 
The increase of the $i^{th}$ node's degree $k_i$ over time can be computed by the probability of a new node connecting to the $i^{th}$ node $\Pi_i$: 

\begin{equation}
\frac{\partial k_i}{\partial t} = m \Pi_i =  m \frac{k_i}{\sum_{j=1}^{N-1} k_j}
\end{equation}

where $m$ is the initial degree of each node upon its introduction at time $t_i$ and $N$ is the total number of nodes in the network at time $t$. Solving the above equation leads to: 

\begin{equation}
k_i(t) = m (\frac{t}{t_i})^\beta
\end{equation}

with $\beta =  \frac{1}{2}$. 
Let $\tau_i = t/t_i$ denote the time factor, the above can be written as: 

\begin{equation}
k_i(t) 	= m \tau_i^{\beta}
\label{eq:kt}
\end{equation}

Given $\beta \in (0,1)$, the $\tau^\beta$ function roughly approximate the schematic trend of cumulative citations over time, as illustrated in Figure~\ref{fig:ct} (b). While there are other citation aging functions such as those proposed by \citet{burrel:cite-age} and \citet{aging}, the scale-free model is very generalizable and has produced results consistent with many real world networks in power-law frequency distributions. 
For this reason, we adopt the functional form of $\tau^\beta$ in the proposed model, where $\beta$ is to be estimated. Although this is not necessarily the most precise method to model citation growth, it does capture the decaying pattern of citations over time, as shown in Figure \ref{fig:ct} (b). 

\subsection{Fitness Modeling}

So far, the model in Equation~\ref{eq:kt} is solely based on the time factor $\tau$, similar to the original scale free model. Incorporation of factors related to individual nodes' competitiveness has led to new formulations such as the fitness model, in which younger nodes with a higher fitness parameter can overcome the dominance of early players \citep{fitness}. We reason that a paper's {\it fitness}, in the citation analysis context, is associated with scholarly establishment and scientific merit of that written communication. This represents a dimension independent of time and its impact on citations manifests over time. Similar to parameter $m$ in Equation~\ref{eq:kt}, the {\it fitness} factor potentiates a node's ability to attract citations and can be measured at the moment of its introduction. Different from $m$, however, {\it fitness} represents individual competitiveness and varies from one node to another. By introducing fitness factor $\eta_i$ of node $i$ to Equation~\ref{eq:kt}, we obtain the following fitness model for citation analysis: 

\begin{equation}
k_i(t) 	\propto \eta_i \tau_i^\beta
\label{eq:kt2}
\end{equation}

where $\eta_i$ represents a collection of factors associated with a node's competitiveness in the citation network and can be further factorized by these variables. While \cite{science:rdnnet} obtained $\beta = 1/2$ in the scale-free model, we leave this to empirical validation in the data. 

Suppose a number of factors contribute to a paper's fitness. Assume $n$ factors, $[\phi_{1i},\phi_{2i},..,\phi_{n,i}]$, can be measured in the data while others are unknown and denoted as $\epsilon_i$. Examples of $\phi$ variables include evidence about existing influences of a paper's authors, e.g., how frequently the authors have been cited prior to the paper's publication. These factors in a sense represent the introductory degree of a node, similar in spirit to parameter $m$ in Equation~\ref{eq:kt}. 

Seen in the light of a power-law degree distribution, there is a huge divide in connectivity between highly cited nodes and those that are rarely connected to. To integrate these degree-related factors in $\eta$ requires normalizing their values (citation scores) from magnitudes' differences to a reasonable scale. Log transformation appears to be a reasonable step in the modeling. We propose: 

\begin{eqnarray}
\ln{\eta_i} = c + \sum_n \gamma_n \ln{\phi_{n,i}} + \delta \ln{\epsilon_i}
\label{eq:f}
\end{eqnarray}

where $\gamma_n$ and $\delta$ represent weights of the contributing factors. 

Equation~\ref{eq:f} is equivalent to: 

\begin{eqnarray}
\eta_i \propto \big( \prod_n \phi_{n,i}^{\gamma_n} \big) \epsilon_i^\delta
\label{eq:f2}
\end{eqnarray}

Replacing $\eta_i$ with Equation~\ref{eq:f2} in Equation~\ref{eq:kt2}, we get the final {\it fitness} model: 

\begin{eqnarray}
k_i(t) 		& = & \alpha \eta_i \tau_i^\beta \nonumber \\
			& = & \alpha \big( \prod_n \phi_{n,i}^{\gamma_n} \big) \tau_i^\beta \epsilon_i^{\delta}
\label{eq:kft}
\end{eqnarray}

where $\tau$ is the time factor and $\phi$ denotes factors about a paper's fitness in terms of existing influences. The coefficients $\alpha$, $\beta$, and $\gamma_n$ can be estimated from data. Let $\alpha' = \ln{\alpha}$ and $\epsilon'_i = \delta \ln{\epsilon_i}$. The above model is equivalent to the following equation after logarithmic transformation, for which generalized linear regression can be performed to estimate the coefficients. 

\begin{eqnarray}
\ln{k_i(t)} = \alpha' + \sum_n \gamma_n \ln{\phi_{n,i}} + \beta \ln{\tau_i} + \delta \epsilon'_i
\label{eq:kft2}
\end{eqnarray}

\subsection{Model Implications}

Based on the model presented in Equation~\ref{eq:kft}, or equivalently Equation~\ref{eq:kft2}, we can quantify proportionally various factors' contributions to a paper's scholarly impact (citations). We have taken into account three categories of variables,  namely $\tau$ the time factor, $\phi$ variables about established influences prior to a paper's publication which are potentially measurable, and other unknown variables contributing to a paper's {\it fitness} summarized as $\epsilon$. 

By singling out the individual variables and estimating related coefficients from data, the model supports evaluation of scholarly impact at multiple levels. For example, isolating the impact of time factor $\tau$ will enable examination of a paper's {\it fitness} and fair comparison of papers regardless of their {\it ages}. In addition, suppose in the data analysis we can include in $\phi$ exhaustive variables about established influences prior to a paper's publication (e.g., authors' prior impacts), then the $\epsilon$ variable is a surrogate of remaining factors about a paper's actual fitness. In this case, quantifying $\epsilon$ will offer insight into a paper's {\it own} ability to attract citations because of its scientific merit and contribution to the field, rather than due to other prior, external factors. 

Finally, we observe that the proposed model has the potential for {\it causality} analysis or prediction of citation scores. In Equation~\ref{eq:kft}, there is a time sequence from right (independent variables) to left (dependent variable). Besides the $\tau$ factor, all variables are about factors prior to or upon the publication of a paper. They can be measured at the time of publication. A citation score $k_i$ can then be seen as the result of these factors over the course of time $\tau$. Because of this time sequence, it is plausible -- though not in definitive terms -- to tell a causal relationship between predictor variables $\phi$ and ultimate citation scores.

\section{Model Validation and Data Analysis}

We apply the proposed fitness model to a collection of 31 years' citation records in information visualization (InfoVis) to validate the model and to analyze evolving patterns about the domain. 
In this section we describe the data, related variables used in the fitness model for paper citations, and a derived model for scholars in the analysis. We discuss results and insights from the analysis in the next section. 

\subsection{Data}

The InfoVis data set is a collection of major publications in the emerging information visualization field during 1974 - 2004 retrieved from the ACM Digital Libraries. It was prepared by the IEEE Information Visualization Contest in 2004 to depict the early history of the field and made available as part of the InfoVis Benchmark Repository \citep{ke:infovis,iv:rep}. 

\begin{figure}[hbt]
\epsfig{file=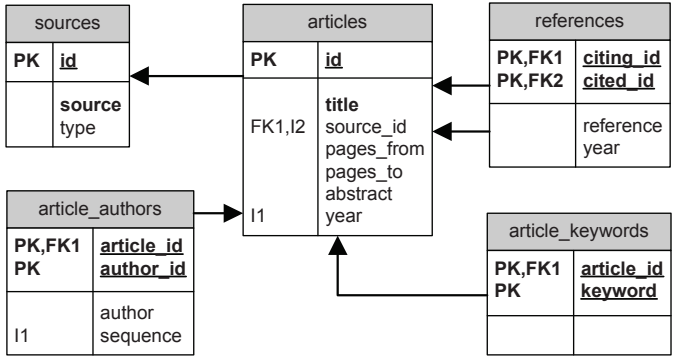,width=3.5in}
\caption{Data schema of the InfoVis 2004 collection}
\label{fig:data1}
\end{figure}

According to \citet{infovis:contest}, the data set contains meta data of important publications on information visualization collected from multiple venues and is representative of the early development (emergence) of the field. The original data is in the XML (Extensible Markup Language) format, which we convert to a relational database. Figure~\ref{fig:data1} shows the database schema with major tables and relationships. 

Each citation record has information such as a paper's title, authors, abstract, keywords, source, references, number of pages, and the year of publication. One paper (acm673478) has no author and is removed from the data. We perform author name unification through automatic name normalization and manual correction. The final data set contains 613 papers with 1,036 unique authors/scholars and 8,502 references to papers within and without the set. 

\begin{figure}[hbt]
\epsfig{file=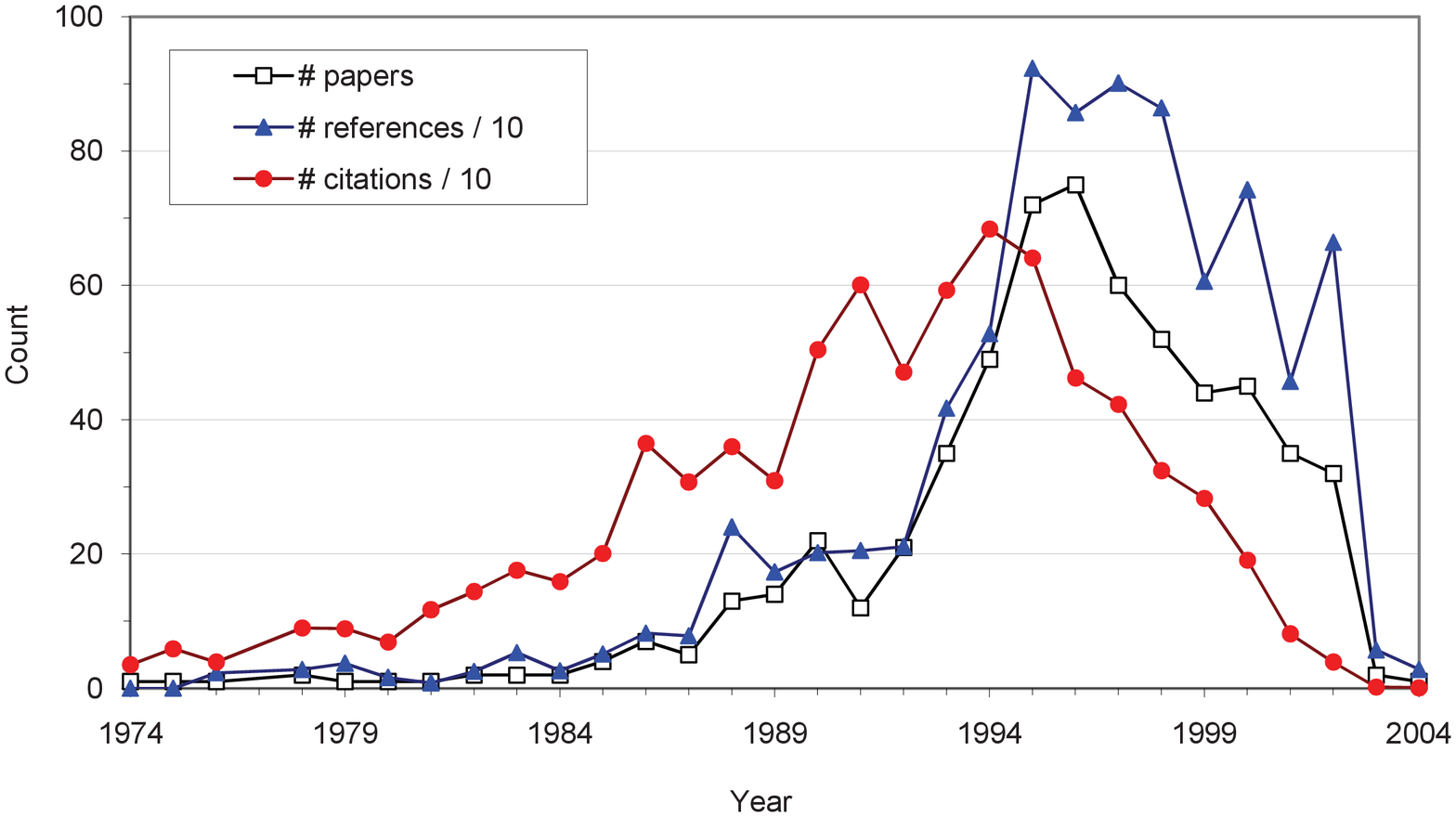,width=4in}
\caption{Yearly distributions of the InfoVis 2004 collection. The number of references refers to the number of references contained in papers published in a year, or the number times other works are referenced by papers published in that particular year, divided by 10 to fit the plot. The number of citations denotes the number of times papers published in a year are cited in following years, divided by 10 to fit the plot.}
\label{fig:data2}
\end{figure}

Yearly distributions of the number of publications, the number of references, and the number of citations are shown in Figure~\ref{fig:data2}. Because the focus of this study is on scholarly impact and {\it fitness} factors based on citation scores, data fields related to content such as title and abstract are not used in the modeling. 

\subsection{Fitness model for papers and related variables}

In terms of the fitness model in Equation~\ref{eq:kft}, we identify related variables from the InfoVis data for model validation and data analysis. We use generalized linear regression in the equivalent form of Equation~\ref{eq:kft2} to estimate related parameters and to examine influences of identified variables. We use $k_i$ to denote the number of citations a paper $i$ received within the collection. Time $\tau_i$ represents the age of paper $i$ when data were collected in 2004. The InfoVis data also have evidence about existing impacts prior to a paper's publication, denoted as $\phi$ variables: 

\begin{enumerate}
  \item Authors' prior impact factor $\phi_{a,i}$ refers to the number of citations authors of paper $i$ received (for earlier works)  before the paper's publication. 
  \item Venue's prior impact factor $\phi_{v,i}$ is the average number of citations to (earlier) papers at the venue where paper $i$ was published before its publication. 
  \item References' prior impact factor $\phi_{r,i}$ denotes the number of citations to works referenced by paper $i$ prior to its publication. 
\end{enumerate}

Note that all $\phi$ variables rely solely on records in the InfoVis collection. We do not seek additional citation information about the collected papers from external sources. Replication of the analysis reported in this article is straightforward. It can be conducted on many other domains where representative citation records in a given time period are available . 


\subsection{Derived model for scholars}

The proposed model has so far focused on the fitness of nodes in citation analysis, where {\it nodes} represent {\it papers}. For an analysis from the perspective of {\it scholars} (authors), a second model can be derived from Equation~\ref{eq:kft}. We treat an author's citation score from a paper as a fair share of the authorship. Using {\it fractional count}, we distribute the credit for a multi-authored work equally among its contributors \citep{coauth:citations,collabProd}. That is, the citation score of each scholar (author) $j$ of paper $i$ is $k_{s,j} = k_i/c_i$, where $k_i$ is citation frequency of paper $i$ and $c_i$ is the number of contributors (authors) of the paper. 

One additional factor for modeling a scholar's citation frequency is the number of papers he or she has authored, denoted as $\rho$. This being considered, factors about authored papers' fitness need to be normalized (averaged) so that $\rho$ is not redundant to existing contributions in individual papers. We propose the use of geometric mean to average contributing variables $\tau$ and $\phi$ for each author. For example, given a set of values $[v_1,v_2,...,v_{\rho_j}]$ for variable $v$ observed in $\rho_j$ papers authored by scholar $j$, the geometric mean $\bar{v}_j$ is computed by: 

\begin{equation}
\bar{v}_j = (\prod^{\rho_j}_{i=1} v_i)^{1/\rho_j} = \sqrt[\rho_j]{v_1 \cdot v_2 \cdot \cdot \cdot v_{\rho_j}}
\label{eq:geomean}
\end{equation}

While arithmetic mean is the common approach to averaging citation scores, research has also adopted harmonic mean and geometric mean in citation analysis \citep{harmean,geomean}. Advantages of geometric mean include reduced standard variance and model simplicity when variables are log-transformable. Using geometric means and additional variable $\rho_j$, the fitness model for scholar $j$ according to Equation~\ref{eq:kft} can be written as:

		\begin{eqnarray}
			k_{s,j} 	& = & \alpha \Big( \prod_n \bar{\phi}_{n,j}^{\gamma_n} \Big) \bar{\tau}_j^\beta \rho_j^\kappa \epsilon_j^{\delta} \\
					& = & \alpha \Big( \prod_n \big((\prod_{i=1}^{\rho_j} \phi_{n,i}^{\gamma_n})^{1/\rho_j} \big)^{\gamma_n} \Big) \big(\prod_{i=1}^{\rho_j} \tau_i^\beta\big)^{1/\rho_j} \rho_j^\kappa \epsilon_j^{\delta} \\
					& = & \alpha \Big( \prod_{i=1}^{\rho_j} \big( \underbrace{(\prod_n \phi_{n,i}^{\gamma_n}) \tau_i^\beta}_{\propto k_i} \big) \Big)^{1/\rho_j} \rho_j^\kappa \epsilon_j^{\delta}
		\label{eq:scholars}
		\end{eqnarray}

Here $(\prod_n \phi_{n,i}^{\gamma_n}) \tau_i^\beta$ turns out to be part of the fitness model for paper $i$ in Equation~\ref{eq:kft}. By using geometric means, the two models for papers and scholars are tightly associated. This derived fitness model for scholars, shown in Equation~\ref{eq:scholars}, can be seen as aggregation of normalized individual papers' contributions toward author citations.

\section{Results}

We present model validation and analysis results from modeling the InfoVis data. We focus on the fitness model for papers and discuss results from the derived model for scholars (authors) as well. We also present results from additional analyses of frequency distributions and multi-authorship impacts. 

\subsection{Model validation}

\subsubsection{Fitness model for papers}

Based on the fitness model for papers in Equation~\ref{eq:kft}, generalized linear regression of the InfoVis data produces estimates of coefficients in Table~\ref{tab:papers}. As results indicate, all estimates are statistically significant in the InfoVis data, where prior fitness $\phi$ and time $\tau$ variables contribute positively to paper citation frequencies.

\begin{table}[htb]
\caption{Fitness model for papers in the InfoVis data}
\begin{tabularx}{\textwidth}{Xcccrl} \midrule
Coefficient								& Estimate	& Std Error 	& t value	& $p_r$($>$$|t|$) & \\ 
\addlinespace \toprule
$\alpha$								& $-0.771$	& $0.115$	& $-6.72$ & 4.3E-11 & *** \\
$\gamma_a$: author impact $\phi_a$		& $0.326$	& $0.0199$	& $16.3$ & 5.5E-50 & *** \\
$\gamma_v$: venue impact $\phi_v$		& $0.0814$	& $0.0365$	& $2.23$ & 0.026 & * \\
$\gamma_r$: refernece impact $\phi_r$	& $0.0395$	& $0.0135$	& $2.93$ & 0.0035 & ** \\
$\beta$: time factor $\tau$				& $0.573$	& $0.048$	& $11.9$ & 1.1E-29 & *** \\
\midrule
\multicolumn{6}{l}{$R^2=0.473$ (adj. $0.469$), $F=136$ on $4$ and $608$ DF} \\ 
\midrule
\end{tabularx}
\label{tab:papers}
\end{table}

The fitness model for papers based on estimates in Table~\ref{tab:papers} can be expressed as Equation~\ref{eq:res:papers}, where the growth of citations $k$ over time $\tau$ follows the function $\tau^{0.57}$, close to scale-free model derivation \citep{science:rdnnet}. While prior impact factors $\phi$ all contribute to a paper's overall ability to attract citations, authors' prior impact factor $\phi_a$ appears to have a greater impact $\gamma_a=0.326$.

\begin{eqnarray}
k(t) 	& = & e^{-0.771} 
				\cdot \phi_a^{0.326} \cdot \phi_v^{0.0814} \cdot \phi_r^{0.0395} 
				\cdot \tau^{0.573} \cdot \epsilon' \nonumber \\
		& = & _{0.462} 
				\cdot \phi_a^{0.326} \cdot \phi_v^{0.0814} \cdot \phi_r^{0.0395} 
				\cdot \tau^{0.573} \cdot \epsilon'
\label{eq:res:papers}
\end{eqnarray}

The fitness model for papers explains nearly $50\%$ of citation score variances in the InfoVis data ($R^2 = 0.473$ and adjusted $R^2 = 0.469$). Given only four factors included in the model, this is relatively high. For example, \citet{det:citations} studied fourteen determinants of citation scores in the discipline of chemical engineering and their model explained $58\%$ of the variance. Ohter models, with an aim to boost prediction accuracy, involved a wide spectrum of content and bibliometric factors \citep{fu:predict}. 

The proposed model only takes into account external variables such as prior fitness factors and time. Without analysis of inherent characteristics such as paper content and scientific merit, the nearly $0.5$ coefficient of determination is considerably large. This supports the assertion that citation growth is indeed a cumulative advantage process, in which success extensively breeds success \citep{price:bib}. 

\subsubsection{Fitness model for scholars}

Modeling scholar fitness in the InfoVis data based on Equation~\ref{eq:scholars} produces estimates in Table~\ref{tab:scholars}. Authors' prior impact factor $\bar{\phi}_a$ (geometric mean), time factor $\bar{\tau}$ (geometric mean), and the number of papers $\rho$ all have significant impacts on the citation outcome. While venues' prior impacts $\phi_v$ and references' prior impacts $\phi_r$ do not show significant influences on citation outcomes in the data, we reason they contribute positively to citations and their non-significant is likely due to their association with other factors such as prior author impact $\phi_a$.

\begin{table}[htb]
\caption{Fitness model for scholars in the InfoVis data}
\begin{tabularx}{\textwidth}{Xcccrl} \midrule
Coefficient	& Estimate	& Std Error 	& t value	& $p_r(>|t|)$ & \\ 
\addlinespace \toprule
$\alpha$									& $-0.735$	& $0.0845$	& $-8.7$ & 1.3E-17 & *** \\
$\gamma_a$: author impact $\bar{\phi_a}$	& $0.217$	& $0.0154$	& $14.1$ & 1.6E-41 & *** \\
$\gamma_v$: venue impact $\bar{\phi_v}$		& $0.0453$	& $0.0443$	& $1.02$ & 0.31 &  \\
$\gamma_r$: reference impact $\bar{\phi_r}$	& $0.0159$	& $0.00842$	& $1.88$ & 0.06 & . \\
$\beta$: time factor $\bar{\tau}$			& $0.395$	& $0.0303$	& $  13$ & 4E-36 & *** \\
$\kappa$: $\#$ authored papers $\rho$		& $0.786$	& $0.0276$	& $28.5$ & 4.5E-132 & *** \\
\midrule
\multicolumn{6}{l}{$R^2=0.629$ (adj. $0.627$), $F=349$ on $5$ and $1030$ DF} \\
\midrule
\end{tabularx}
\label{tab:scholars}
\end{table}

Given estimates in Table~\ref{tab:scholars}, the fitness model for scholars can be expressed as Equation~\ref{eq:res:scholars}. Citation growth over time follows the rough functional form of $\tau^{0.4}$, to which the scale-free model remains a fine approximation \citep{science:rdnnet}. Apparently, the number of papers a scholar authored $\rho$ has great influences on the citation outcome. The relation between scholar citation frequency $k_s$ and $\rho$ is $k_s \propto \rho^{0.8}$, close to a linear function.

\begin{eqnarray}
k_s(t) 	& = & e^{-0.735} 
				\cdot \phi_a^{0.217} \cdot \phi_v^{0.0453} \cdot \phi_r^{0.0159} 
				\cdot \tau^{0.395} \cdot \rho^{0.786} \cdot \epsilon' \nonumber \\
		& = & _{0.479} 
				\cdot \phi_a^{0.217} \cdot \phi_v^{0.0453} \cdot \phi_r^{0.0159} 
				\cdot \tau^{0.395} \cdot \rho^{0.786} \cdot \epsilon'
\label{eq:res:scholars}
\end{eqnarray}

Again, the analysis indicates significant impacts of {\it preferential attachment} in citation growth as a cumulative advantage process \citep{price:bib,science:rdnnet}. The large $R^2 > 0.6$ from modeling InfoVis scholars suggests that there is an extensive rich-get-richer and fitter-get-richer effect. Scholarly productivity and impact evaluation based on raw citation scores is not necessarily fair given existing advantage of early players and bias caused by scholarly establishment. Isolating the influences of time and prior fitness factors may lead to new insight into the evaluation, which we will discuss later. 

\subsubsection{Validation of citation frequency distributions}

\begin{figure}[htb]
\begin{tabular}{cc}
\begin{minipage}{2in}
\epsfig{file=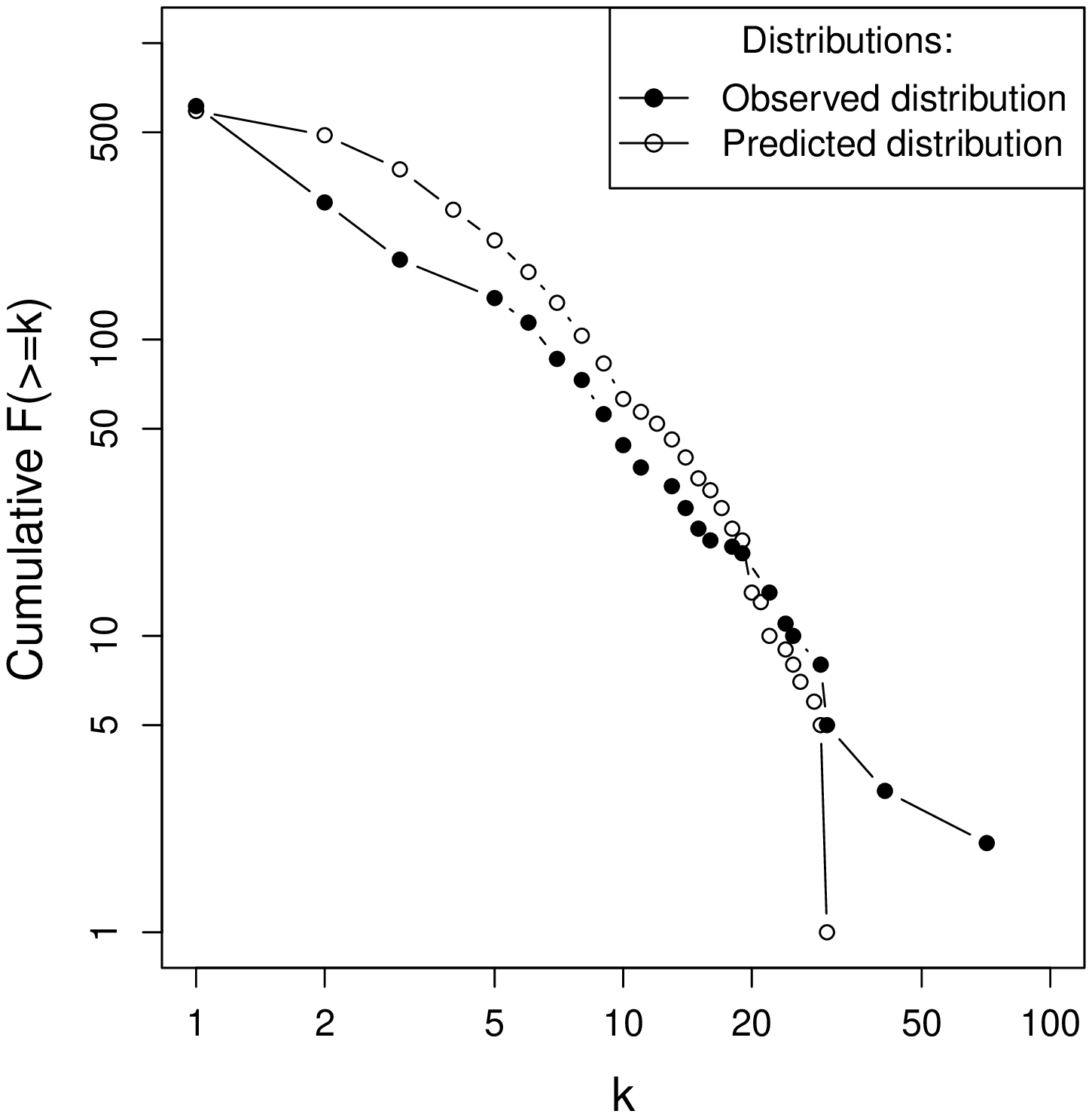,width=2in}
\end{minipage}
&
\begin{minipage}{2in}
\epsfig{file=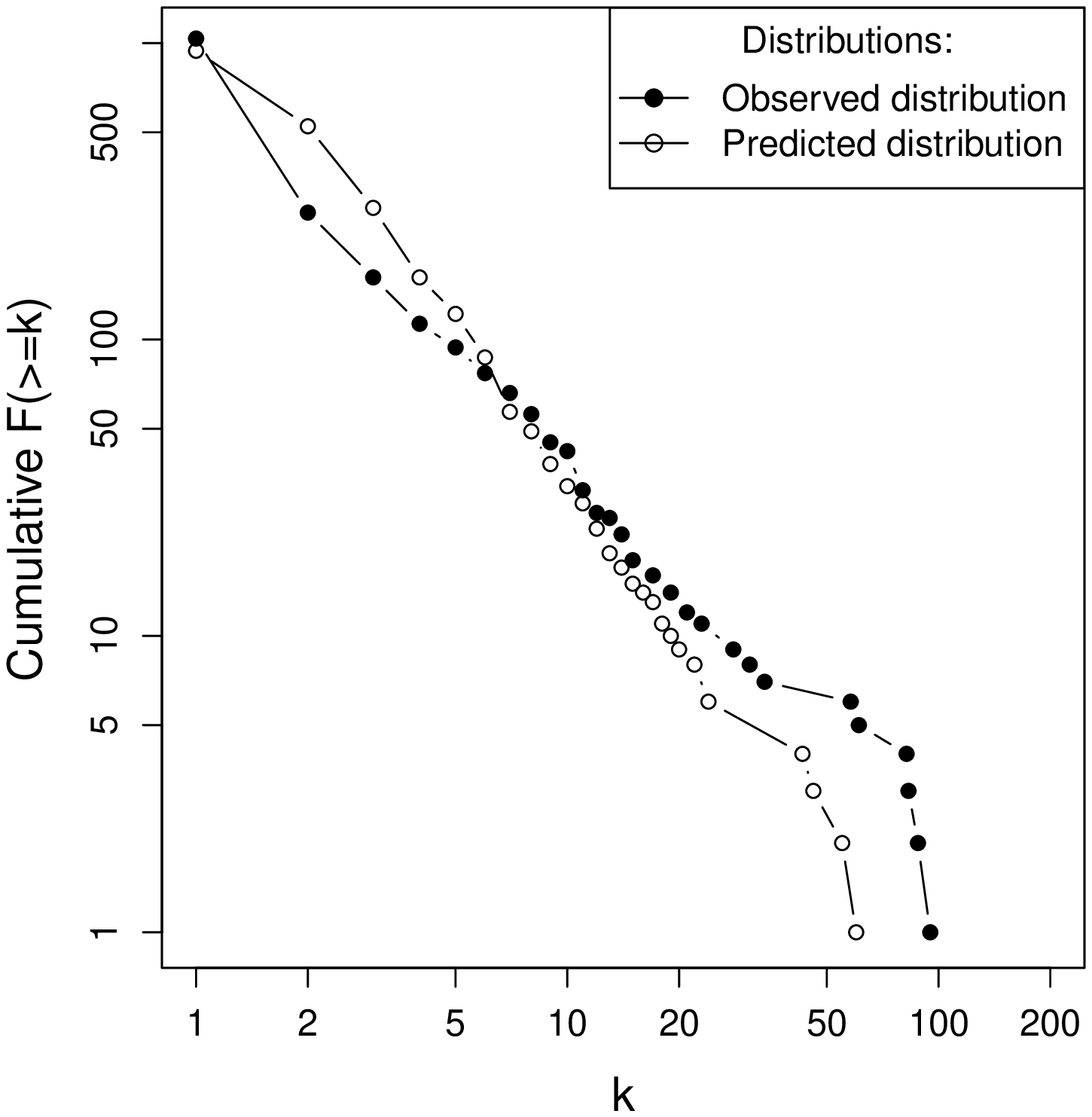,width=2in}
\end{minipage}
\\
(a) papers & (b) scholars
\end{tabular}
\caption{Cumulative citation score distributions. 
Both figures are on log/log coordinates. 
(a) depicts observed and predicted citation distributions of papers. 
(b) shows observed and predicted citation distributions of scholars. 
Data points are weighted proportionally to their citation scores in the fitness models plotted here. 
}
\label{fig:dist}
\end{figure}

Connectivity distribution analysis has been an important tool in network science research. A major goal of various complex network models has been to reproduce important patterns/characteristics in these distributions. Here we use the two models discussed above to generate citations distributions for papers as well as for scholars. Figures~\ref{fig:dist} (a) and (b) show cumulative frequency distributions for the two models respectively and compare their predicted results to observed distributions in the data.

In general, distributions generated by the proposed fitness models manifest cumulative patterns similar to those observed in the data. Predicted citation frequencies appear to be {\it conservative} estimates of observed frequencies. Overall, the models produce more rarely cited nodes (top-left in both figures) and fewer highly cited ones (bottom-right in both figures). For highly-cited nodes that are predicted by the models, their citation frequencies are smaller than actual values (bottom-right in both figures). Despite these local differences, predicted (model-generated) and observed distributions look consistent. 

\subsubsection{Top papers and scholars in InfoVis}

\begin{table}[htb]
\caption{Top rank papers by citation scores $k$ in the data. 
Column $k_t$ has citation scores normalized by time factor $\tau$, that is $k_t = k / \tau^\beta$. 
Column $k_{tf}$ denotes citation scores normalized by time factor $\tau$ and $\phi$ variables, that is $k_{tf} = k / (\tau^\beta \prod_n \phi_n^{\gamma_n})$.
Column $k_{acm}$ is the number of citations identified for each paper in the ACM DL in 2012. 
}
\begin{tabularx}{\textwidth}{lXcrrrr} \midrule
No. & Title & Year & $k$ & $k_t$ & $k_{tf}$ & $k_{acm}$ \\
\addlinespace \toprule
1	& Cone Trees: animated 3D...	& 1991	& 70	& 33.9	& 6.4	& 300\\
2	& The perspective wall: detail and...	& 1991	& 30	& 14.8	& 2.7	& 200\\
3	& Visual information seeking: tight...	& 1994	& 29	& 16.4	& 3.2	& 230\\
4	& Information visualization using 3D...	& 1993	& 28	& 15.1	& 2.5	& 142\\
5	& Tree-Maps a space-filling approach to 	& 1991	& 28	& 13.8	& 2.8	& 212\\
6	& The table lens: merging graphical and 	& 1994	& 24	& 13.7	& 2.7	& 151\\
7	& Pad++: a zooming graphical interface...	& 1994	& 23	& 13.1	& 4.5	& 202\\
8	& Pad: an alternative approach to the...	& 1993	& 22	& 12.0	& 6.7	& 151\\
9	& Stretching the rubber sheet: a...	& 1993	& 22	& 12.0	& 3.7	& 79\\
10	& Dynamic queries for information...	& 1992	& 19	&  9.9	& 2.1	& 127\\
11	& A review and taxonomy of distortion-or	& 1994	& 19	& 10.9	& 3.8	& 120\\
12	& Tree visualization with tree-maps: 2-d	& 1992	& 18	&  9.5	& 2.3	& 215\\
13	& Graphical Fisheye Views	& 1993	& 16	& 8.9	& 2.6	& 122\\
14	& Toolglass and magic lenses: the see-th	& 1993	& 15	& 8.3	& 3.3	& 346\\
15	& Parallel coordinates a tool for...	& 1990	& 15	& 7.3	& 2.8	& 145\\
16	& The movable filter as a user...	& 1994	& 13	& 7.7	& 2.5	& 72\\
17	& Worlds within worlds: metaphors for...	& 1990	& 13	& 6.4	& 2.7	& 54\\
18	& The dynamic HomeFinder: evaluating...	& 1992	& 12	& 6.5	& 1.5	& 75\\
19	& Interactive graphic design using...	& 1994	& 12	& 7.1	& 2.0	& 49\\
20	& To see, or not to see- is That the...	& 1991	& 10	& 5.2	& 3.0	& 24\\\midrule
\multicolumn{3}{r}{Correlation with $k_{acm}$:}&0.586& {\bf 0.593} &0.394&  \\ \midrule\end{tabularx}
\label{tab:toppapers}
\end{table}

With treatments on time and fitness factors, the proposed model has the potential to single out these variables and to identify an individual node's ability for long-term growth. The InfoVis data were prepared in 2004 to document the birth and early history of the field. Now that many years have past, there is new evidence about how well nodes (papers and scholars) have grown in citations. We use the total number of citations identified for each paper or scholar in the ACM digital libraries in 2012, denoted as $k_{acm}$, as a surrogate of its long-term impact. 

We sort papers by their overall citation scores $k$ within the InfoVis 2004 data and select the top $20$ for Table~\ref{tab:toppapers}. We reason that removing the time factor $\tau$ from citation scores, among others, supports a fairer comparison of papers published in various years. This leads to weighted citation scores $k_t$ based on time normalization $k_t = k / \tau^\beta$ and further reduction of fitness factors in $k_{tf} = k / (\tau^\beta \prod_n \phi_n^{\gamma_n})$ (see additional $k_t$ and $k_{tf}$ columns in Table~\ref{tab:toppapers}). While $k$ and $k_t$ are in general consistent with the $k_{acm}$ outcome, $k_t$ appears to have a slightly higher correlation with $k_{acm}$ whereas $k_{ft}$, with removal of both time and prior fitness factors, has a weaker correlation. 

\begin{table}[htb]
\caption{Top rank scholars by citation scores $k_s$ in the data. 
Column $k_t$ has citation scores normalized by time factor $\tau$, that is $k_t = k /\bar{\tau}^\beta$; 
whereas $k_{tf}$ denotes citation scores normalized by time factor $\tau$ and $\phi$ variables, that is $k_{tf} = k / (\bar{\tau}^\beta \prod_n \bar{\phi}_n^{\gamma_n})$.
Column $k_{acm}$ is the number of citations identified for each scholar in the ACM DL in 2012. 
}
\begin{tabularx}{\textwidth}{lXrrrr} \midrule
No. & Scholar name (mean pub year) & $k_s$ & $k_t$ & $k_{tf}$ & $k_{acm}$ \\ 
\addlinespace \toprule
1	& B. Shneiderman (1995) 	& 94	& 82.5	& 31.7	& 5389\\
2	& J. D. Mackinlay (1995) 	& 87	& 77.2	& 28.2	& 1736\\
3	& S. K. Card (1995) 	& 82	& 70.3	& 24.3	& 3547\\
4	& G. W. Furnas (1994) 	& 81	& 69.2	& 25.7	& 1595\\
5	& G. Robertson (1994) 	& 60	& 50.9	& 17.3	& 2177\\
6	& E. R. Tufte (1988) 	& 58	& 40.3	& 18.7	& 884\\
7	& C. Ahlberg (1994) 	& 33	& 27.8	& 10.6	& 535\\
8	& R. Rao (1995) 	& 30	& 27.5	& 11.1	& 271\\
9	& W. S. Cleveland (1988) 	& 27	& 19.7	&  9.6	& 311\\
10	& T. Munzner (1997) 	& 22	& 21.9	& 11.8	& 318\\
11	& B. Bederson (1998) 	& 22	& 23.2	& 10.4	& 2094\\
12	& S. K. Feiner (1992) 	& 20	& 16.8	&  9.4	& 2656\\
13	& P. Pirolli (1996) 	& 18	& 17.6	& 6.8	& 1396\\
14	& S. G. Eick (1997) 	& 18	& 17.5	& 8.6	& 736\\
15	& B. Johnson (1991) 	& 16	& 13.1	& 5.5	& 241\\
16	& S. F. Roth (1995) 	& 16	& 15.0	& 7.4	& 471\\
17	& J. D. Hollan (1995) 	& 14	& 13.6	& 6.6	& 967\\
18	& M. H. Brown (1993) 	& 14	& 12.1	& 6.1	& 622\\
19	& M. Chalmers (1997) 	& 13	& 13.7	& 8.4	& 594\\
20	& J. Lamping (1994) 	& 13	& 12.0	& 4.9	& 649\\\midrule
\multicolumn{2}{r}{Correlation with $k_{acm}$:}&0.685& {\bf 0.710} & 0.707 &  \\ \midrule\end{tabularx}
\label{tab:topscholars}
\end{table}

Based on the fitness model for scholars, we perform a similar analysis of these variables, namely: 1) the raw citations of a scholar in the data $k_s$, 2) time normalized citations $k_t = k /\bar{\tau}^\beta$, 3) time and prior fitness normalized citations $k_{tf} = k / (\bar{\tau}^\beta \prod_n \bar{\phi}_n^{\gamma_n})$, and 4) citations in ACM DL by 2012. As shown in Table~\ref{tab:topscholars}, both $k_t$ and $k_{tf}$ have stronger correlations with the long-term citation growth $k_{acm}$ than raw citation score $k_s$ does. Note that the normalized score $k_{tf}$ represents unknown fitness factors in the analysis and may be further factorized in future studies. No significance test has been performed on this specific analysis given the small samples. 

\subsubsection{Paper fitness over time}

\begin{figure}[htb]
\begin{tabular}{ccc}
\begin{minipage}{1.5in}
\epsfig{file=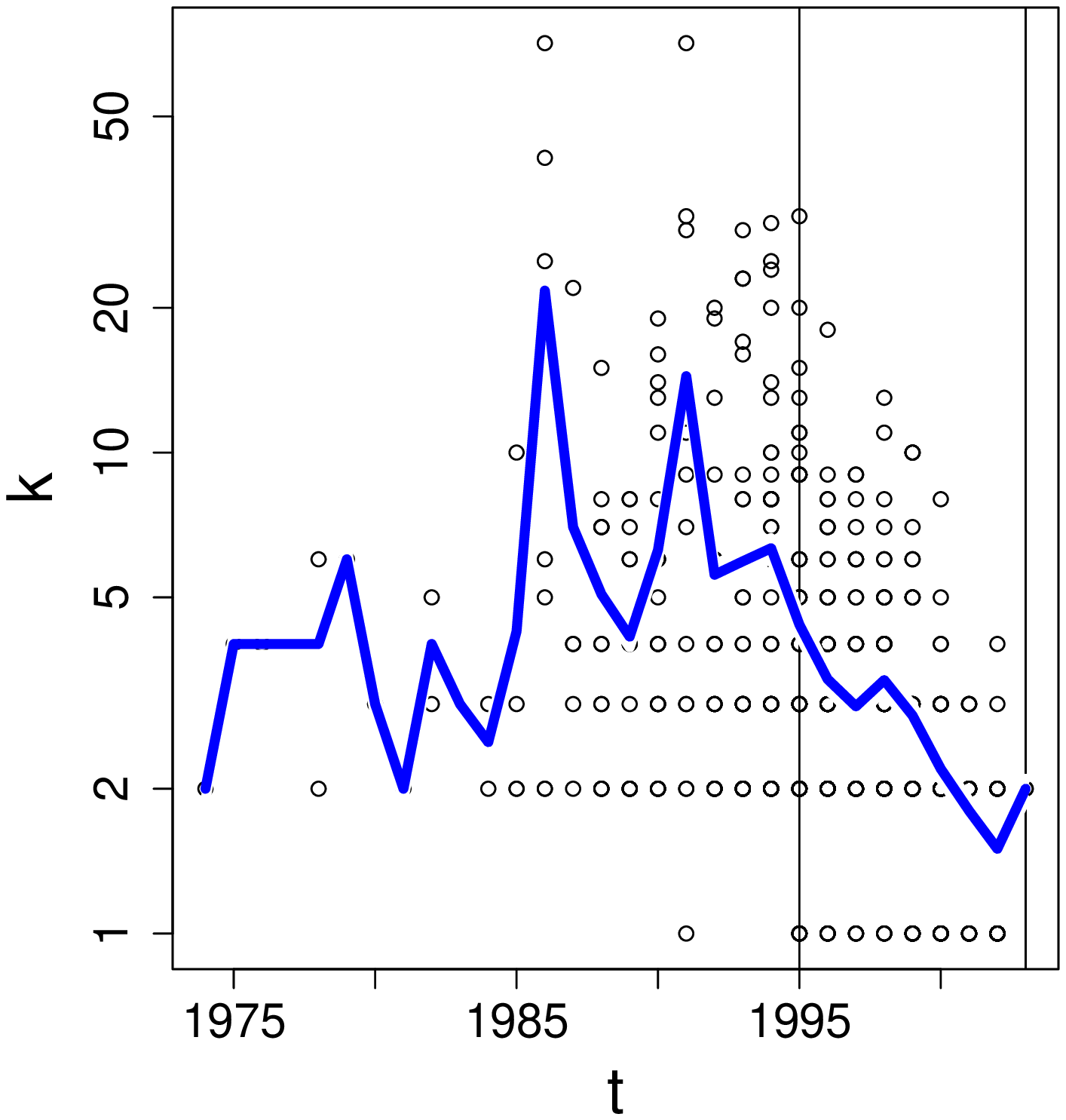,width=1.5in}
\end{minipage}
&
\begin{minipage}{1.5in}
\epsfig{file=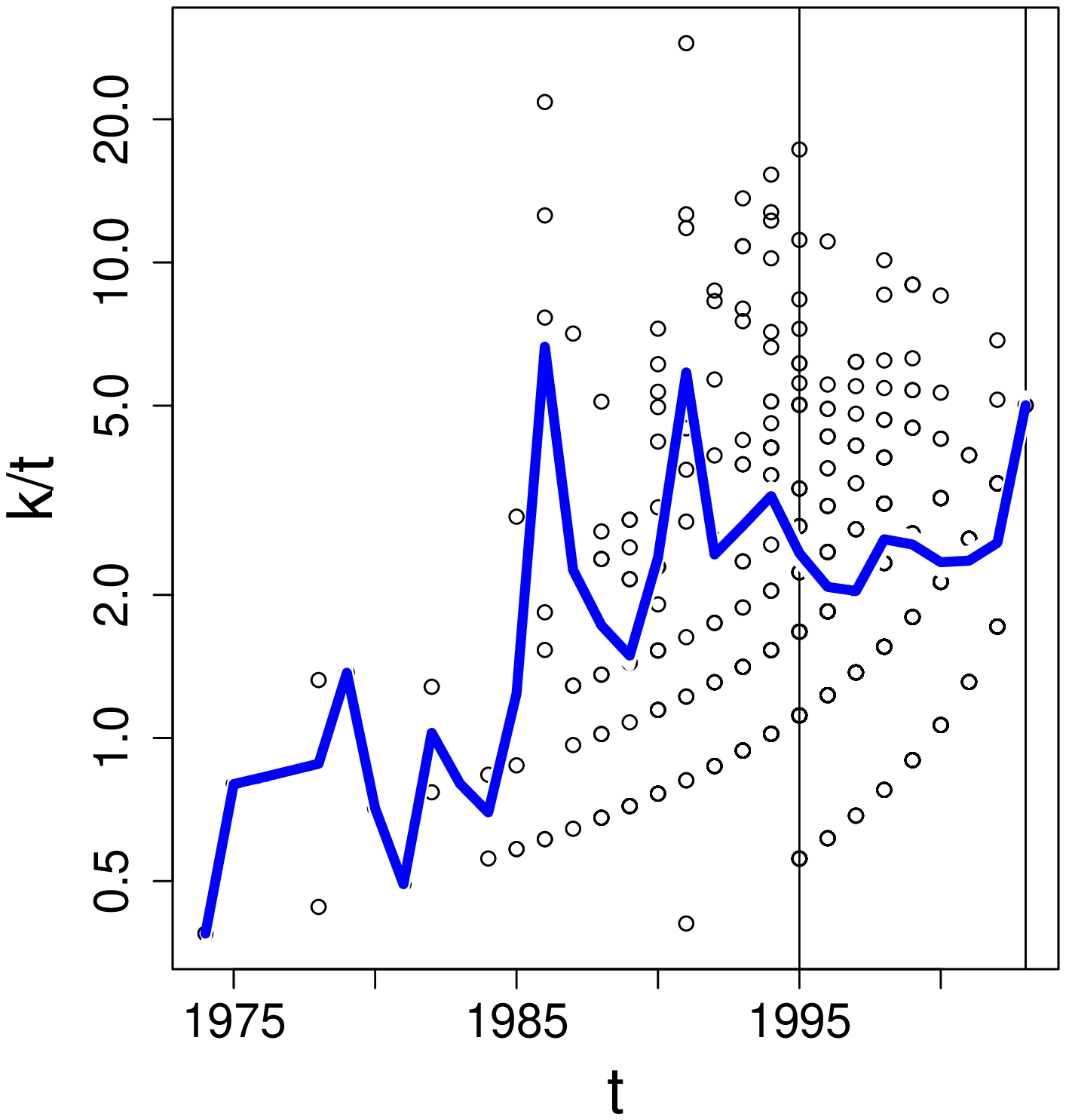,width=1.5in}
\end{minipage}
&
\begin{minipage}{1.5in}
\epsfig{file=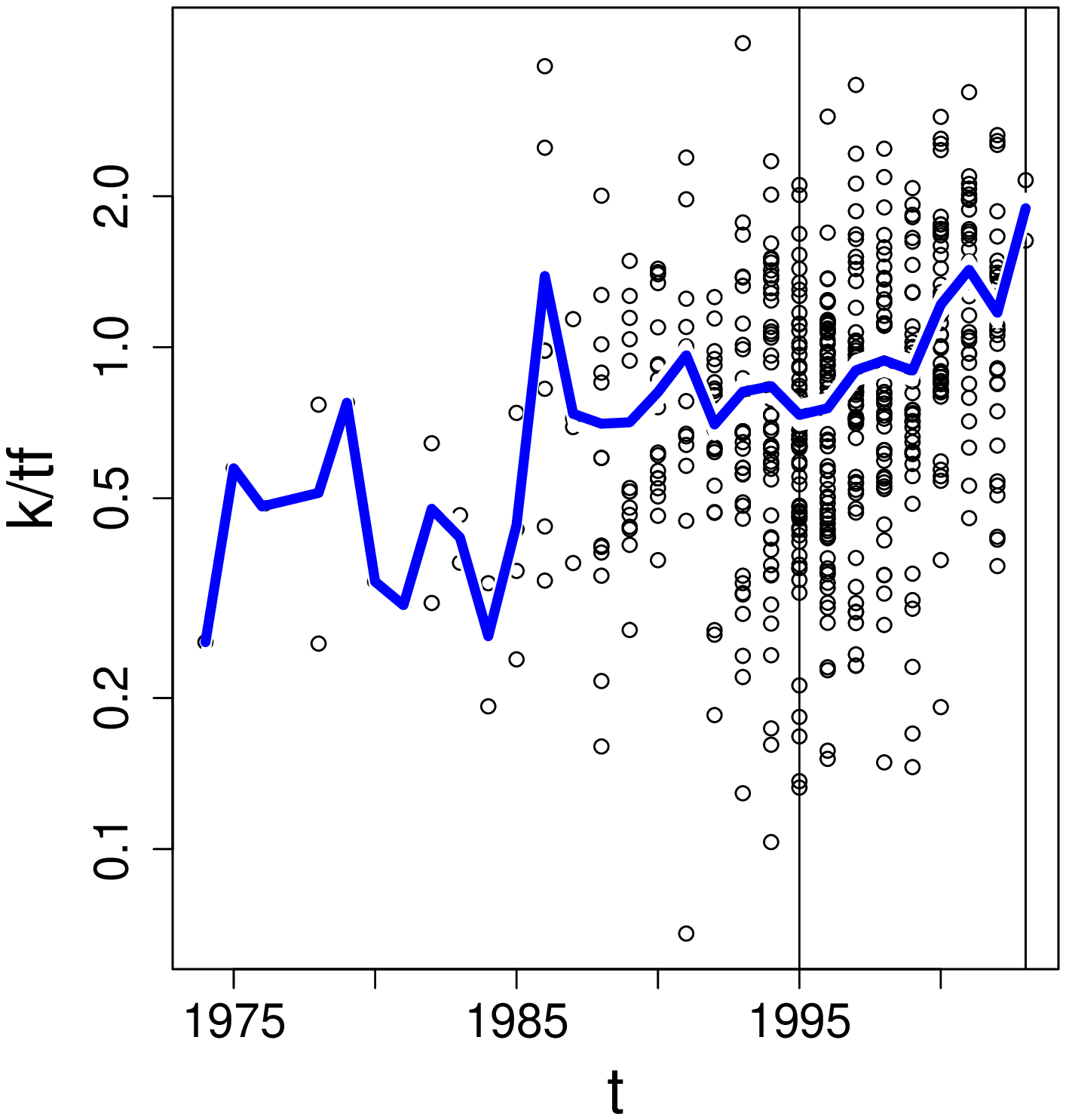,width=1.5in}
\end{minipage}
\\
(a) citations $k$ & (b) $k$ normalized by $\tau$ & (c) $k$ normalized by $\tau$ \& $\phi$
\\
\end{tabular}
\caption{Citation scores over time. 
Each data point represents the citation score $y$ of a paper published in year $x$. 
The solid line (curve) indicates yearly averages. 
Vertical lines show the 9-year period of the InfoVis conference from 1995 - 2003. 
In all figures, $x$ is an ordinary axis and the $y$ axis is logarithmic. 
Figure (a) shows the yearly distribution of raw citation scores 1974 - 2004. 
Figure (b) is the yearly distribution of citation scores normalized by time factor $\tau$, that is $k_t = k / \tau^\beta$. 
Figure (c) is the yearly distribution normalized by time factor $\tau$ and $\phi$ variables, that is $k_{tf} = k / (\tau^\beta \prod_n \phi_n^{\gamma_n})$. 
}
\label{fig:overtime}
\end{figure}

The InfoVis data are about the birth and growth of the scientific field of information visualization. Citation records in the data, though not a exhaustive collection about the field, are generally considered representative of its development. Plotting raw citation scores of papers in the data over time, however, shows a counter-intuitive picture. In Figure~\ref{fig:overtime} (a), average citation score $k$ peaks in late 1980s - early 1990s and decreases continuously afterwards. While we understand that this trend is mainly due to the lack of citation data for recent publications, tracking raw citation scores for analyzing the development of a field does not offer much insight. Note that there is only one paper published in 2004 in the collection, which has no citations and is not included in the plots. 

Plots of the normalized scores $k_t$ and $k_{tf}$ over time, as shown in Figure~\ref{fig:overtime} (b) and (c) respectively, tell a different story. Time normalized $k_t$ also has peaks in late 1980s but moves roughly constantly over time after early 1990s. With $k_{tf}$ normalized by time and prior fitness factors, there appears to be an overall incremental development over the years, especially during 1995 - 2003. According to \citet{infovis:contest}, 1995 marked an important milestone of the field when the first IEEE Symposium on Information Visualization was held. The InfoVis data in this analysis include 9 years' proceedings of the conference from 1995 - 2003. It is quite certain that the field experienced healthy growth during this period and the $k_{tf}$ plot in Figure~\ref{fig:overtime} (c) is relatively consistent with this observation. 

\subsubsection{Summary of model validation and analysis}

The modeling and validation rely on generalized linear regression analyses, citation frequency distributions, comparison with long-term citation evidence, and examination of (normalized) citation frequencies over time. We have found the proposed fitness model to be a useful tool for scholarly impact evaluation, which offers insight consistent with observations about citation growth in general and the InfoVis field in particular. Models produce significant results with the InfoVis data and regenerate citation distribution patterns similar to those in the data. The fitness models, involving time and prior fitness factors, explain a significant portion of citation variances ($R^2 \approx 50\%$ in the model for papers and $R^2 \approx 60\%$ for scholars). Isolation of these factors offers good estimation about nodes' (papers' and scholars') ability to gain citations in the long term. Normalization of citation scores by time and prior fitness factors also leads to a more reasonable depiction of InfoVis development in its recent history. 

\subsection{Additional data analysis}

\subsubsection{Paper fitness distributions}

\begin{figure}[htb]
\begin{tabular}{ccc}
\begin{minipage}{1.4in}
\epsfig{file=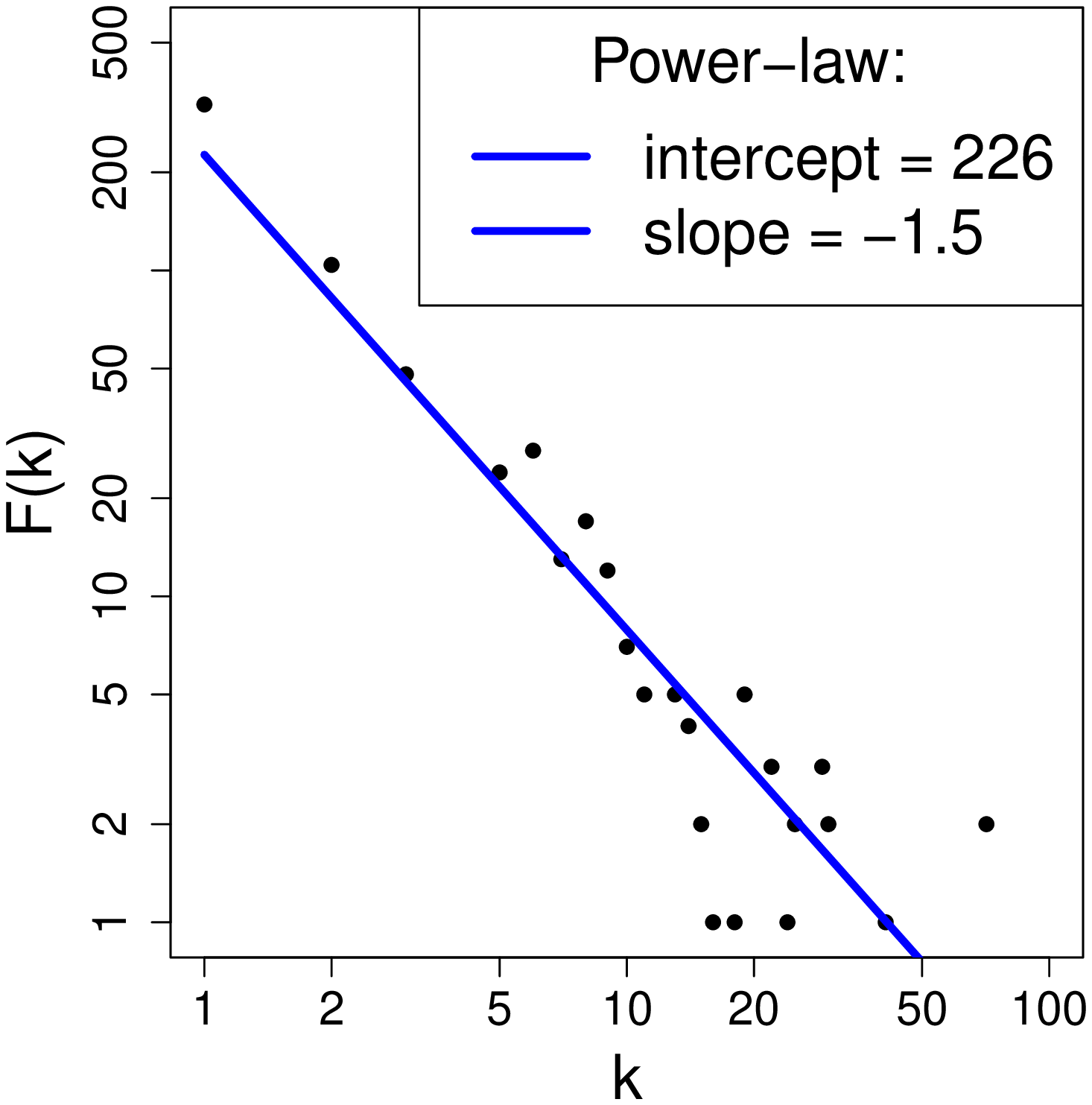,width=1.4in}
\end{minipage}
&
\begin{minipage}{1.4in}
\epsfig{file=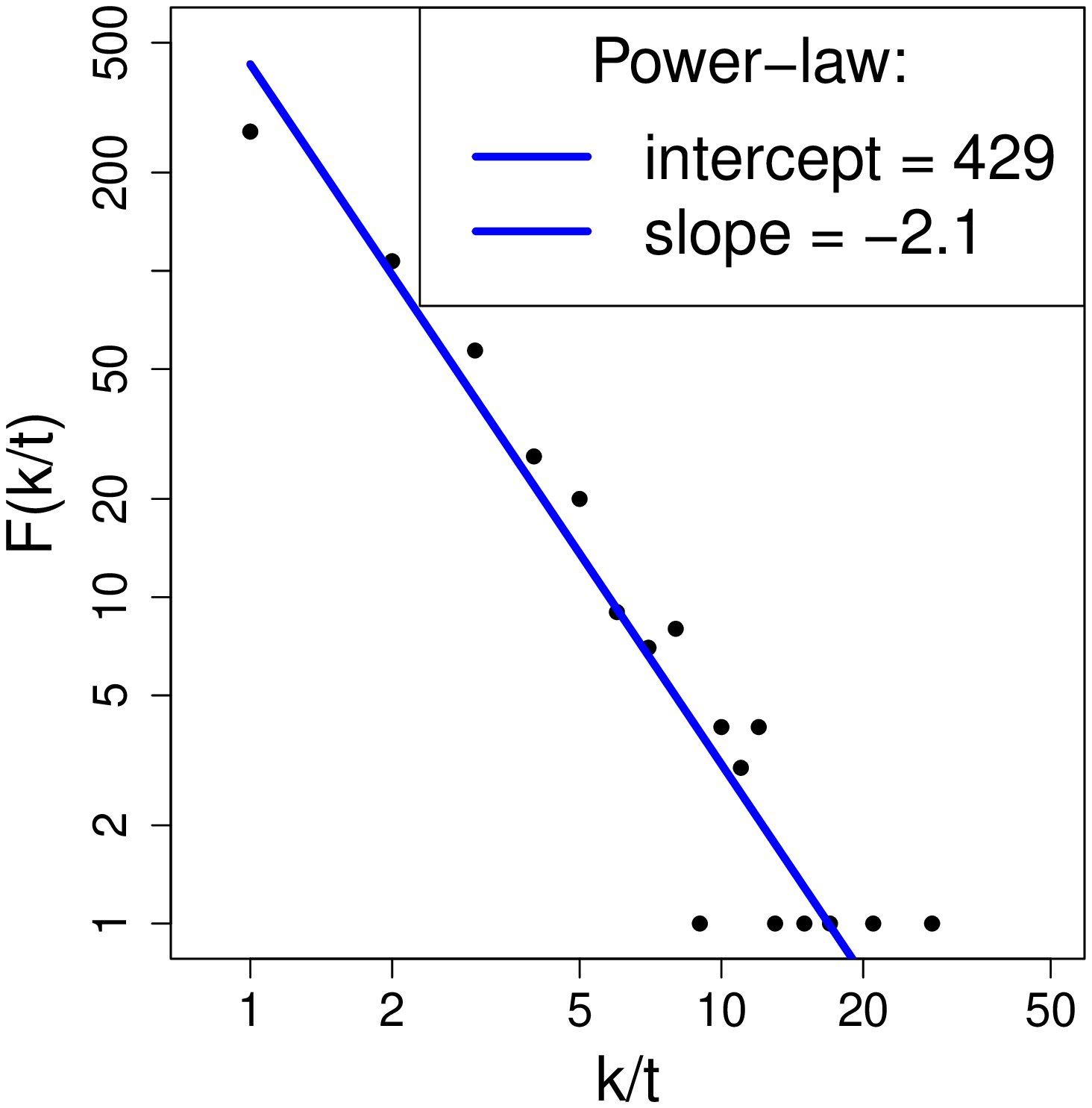,width=1.4in}
\end{minipage}
&
\begin{minipage}{1.4in}
\epsfig{file=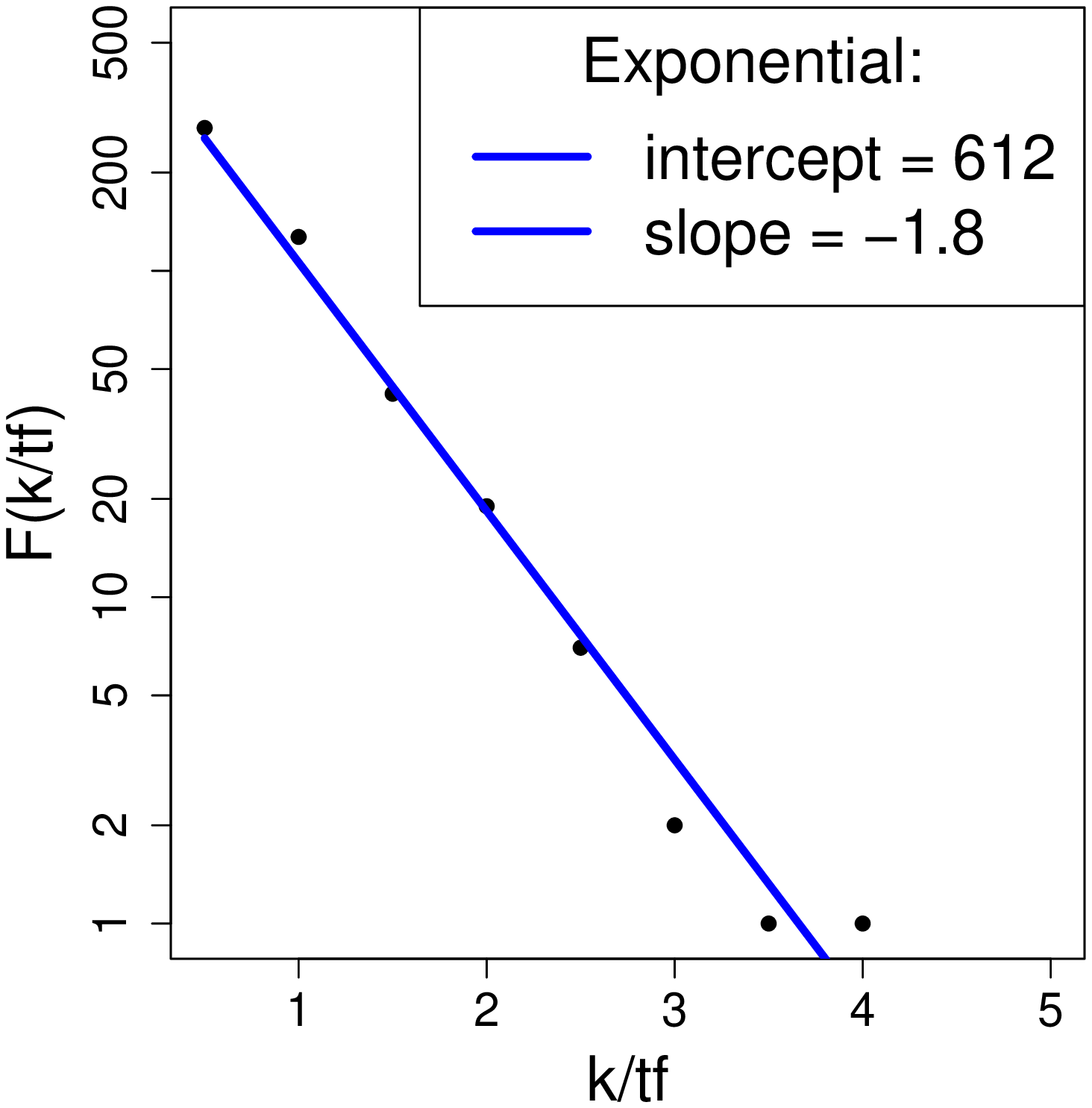,width=1.4in}
\end{minipage}
\\
(a) citations $k$ & (b) $k$ normalized by $\tau$ & (c) $k$ normalized by $\tau$ \& $\phi$
\\
\end{tabular}
\caption{Paper citation score distributions. 
Figure (a) shows the distribution of raw citation scores. 
Figure (b) is the distribution of citation scores normalized by time factor $\tau$, that is $k_t = k / \tau^\beta$. 
Figure (c) is the distribution normalized by time factor $\tau$ and $\phi$ variables, that is $k_{tf} = k / (\tau^\beta \prod_n \phi_n^{\gamma_n})$. 
In figures (a) and (b) $x$ and $y$ coordinates are both logarithmic whereas in figure (c) only $y$ axis is log-transformed. 
}
\label{fig:dist2}
\end{figure}

In Figure~\ref{fig:dist}, we looked at cumulative distributions of raw citation scores. Figure~\ref{fig:dist2} (a) shows the discrete, non-cumulative distribution of paper citation frequencies, which follows a rough power-law function linear on the log-log plot. Figure~\ref{fig:dist2} (b) plots time normalized $k_t$ whereas (c) shows the distribution of $k_{tf}$ with both time and prior fitness normalization.

While the $k_t$ distribution remains roughly linear on log-log (see Figure~\ref{fig:dist2} (b)), the $k_{tf}$ distribution resembles an exponential form (see Figure~\ref{fig:dist2} (c)). The $k_{tf}$ normalization, with the removal of contributions from time and variables related to prior fitness, is essentially an unknown factor about a paper's additional ability to gain citations. It is likely a representative of constituent variables such as a paper's quality, scientific merit, and potential contribution to the field. In this sense, $k_{ft}$ can be seen as a measure about a paper's {\it inherent} fitness whereas prior fitness $\phi$ and time $\tau$ are external factors. The exponential distribution suggests a single- or broad-scale nature of $k_{tf}$. That is there are certain constraints on how related inherent factors may vary, leading to a scale limit \citep{Amaral2000}. Identification of additional variables in data and further factorization of $k_{tf}$ may lead to discovery of important characteristics of this distribution and better understanding of its implications on citation analysis. 

\subsubsection{Impact of multi-authorship}

\begin{figure}[htb]
\epsfig{file=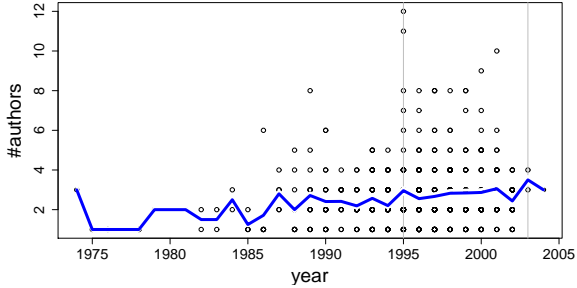,width=3in}
\caption{The number of authors per paper over time. 
Data points represent invidiual papers. 
The solide line indicates yearly averages. 
}
\label{fig:author}
\end{figure}

The average number of authors per paper over time, as shown in Figure~\ref{fig:author}, is generally increasing over 31 years in the InfoVis data. Earlier analysis of the field has shown that collaboration was a key factor in the development of InfoVis given its multidisciplinary nature \citep{borner:globalbrain}. Analyzing multi-authorship contributions will help understand collaboration trends and impacts.

\begin{figure}[htb]
\begin{tabular}{ccc}
\begin{minipage}{1.4in}
\epsfig{file=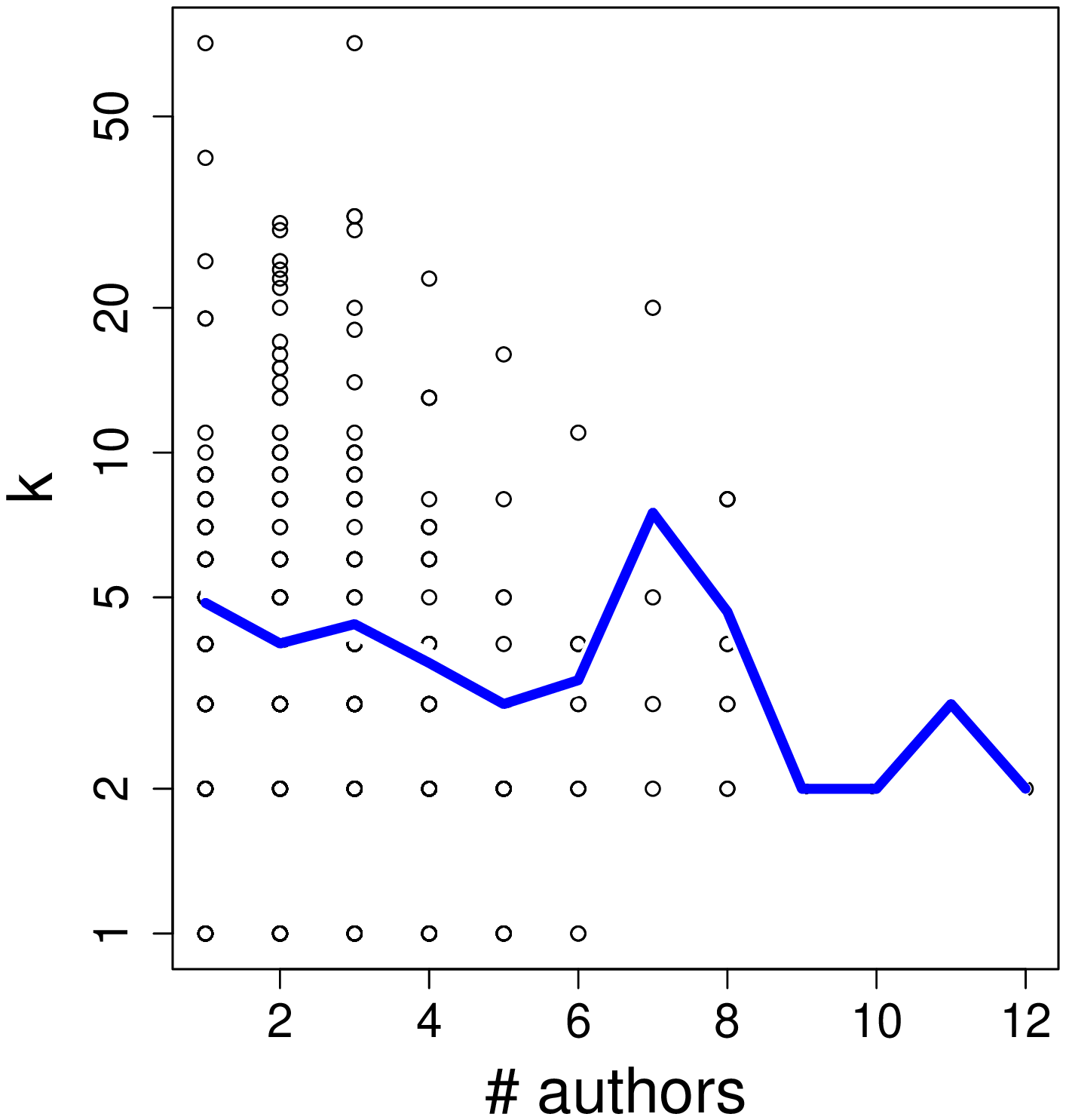,width=1.4in}
\end{minipage}
&
\begin{minipage}{1.4in}
\epsfig{file=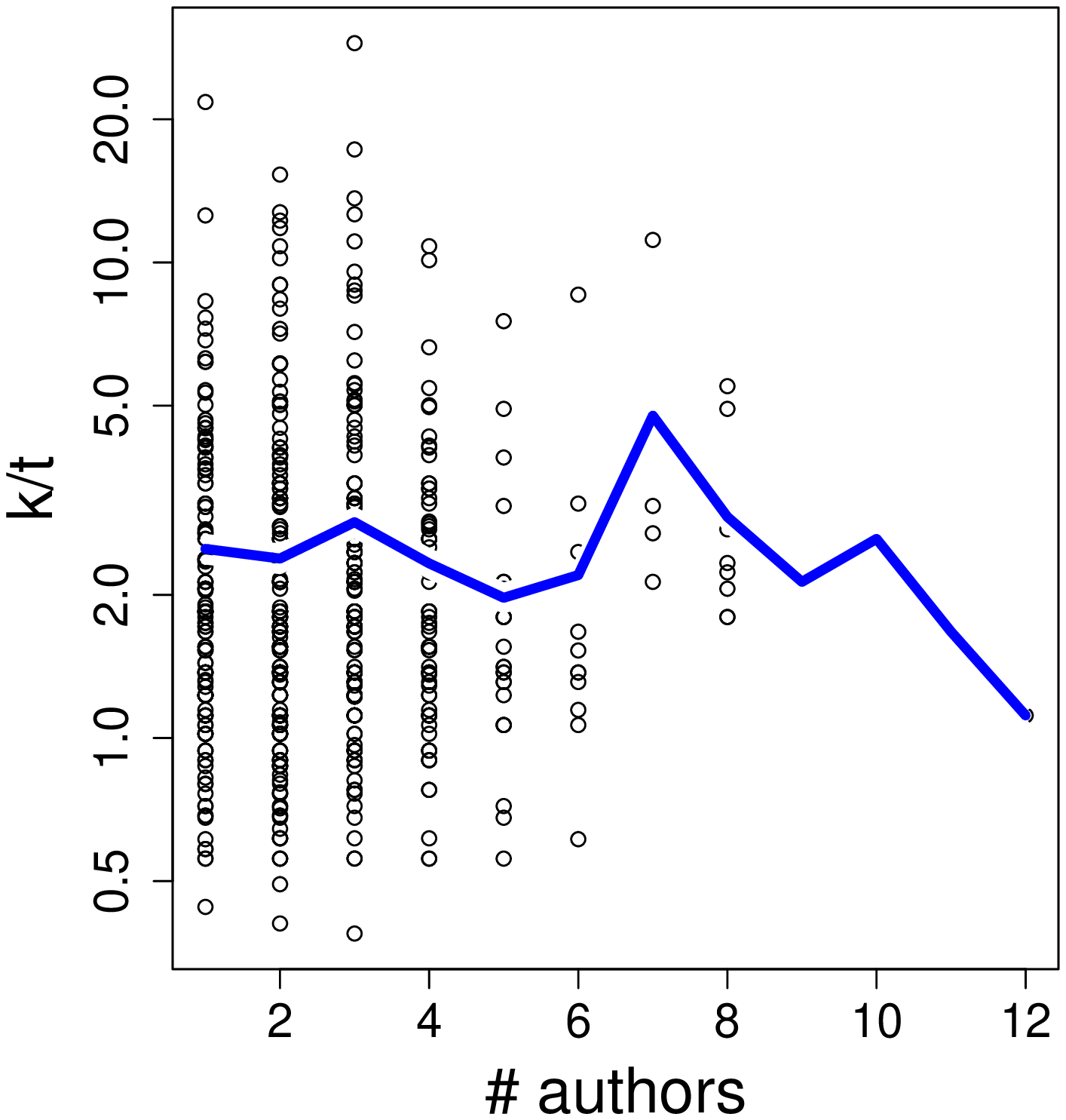,width=1.4in}
\end{minipage}
&
\begin{minipage}{1.4in}
\epsfig{file=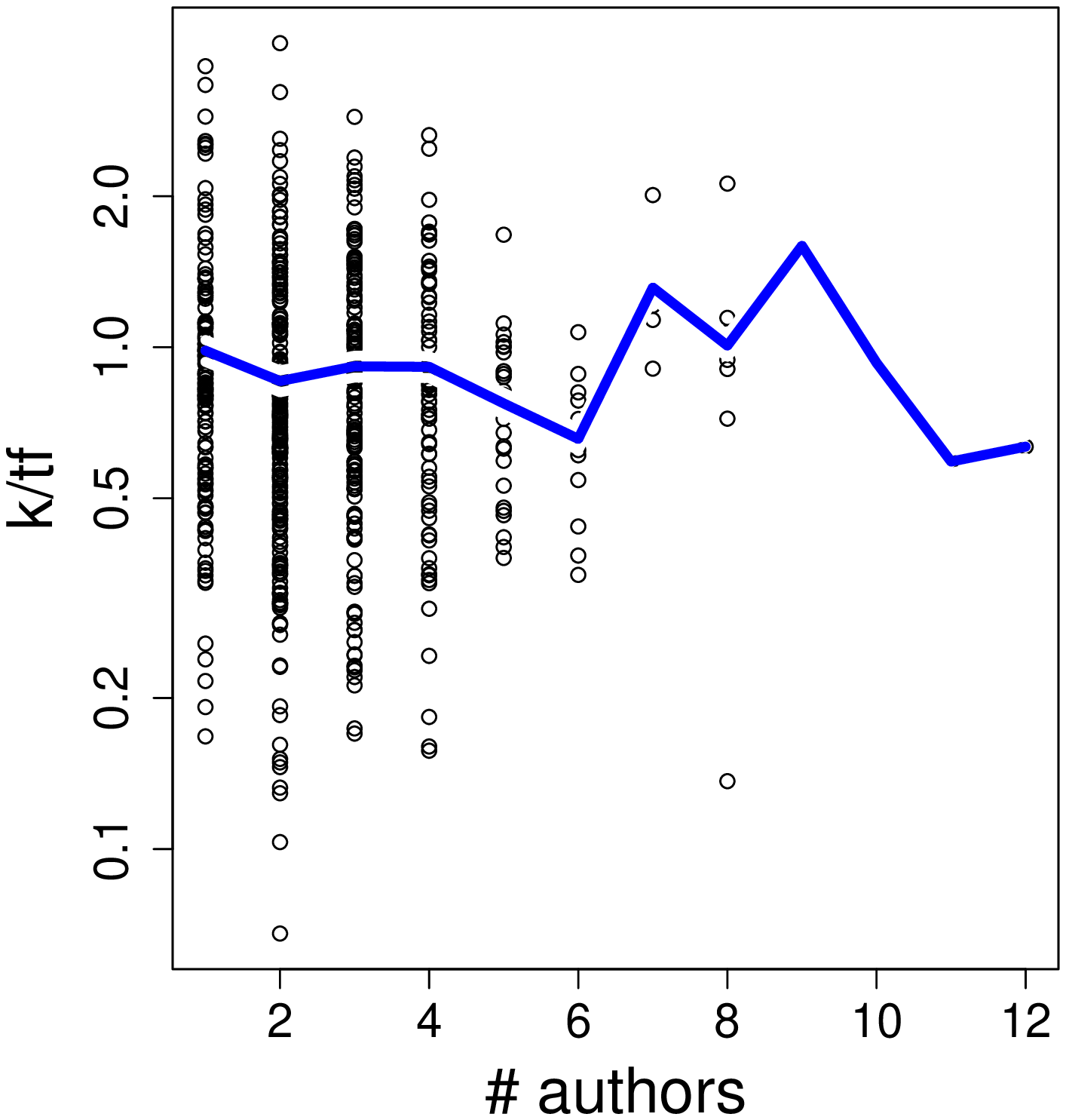,width=1.4in} 
\end{minipage}
\\
(a) citations $k$ & (b) $k$ normalized by $\tau$ & (c) $k$ normalized by $\tau$ \& $\phi$
\\
\end{tabular}
\caption{Citation scores over \# authors. 
Each data point represents the citation score $y$ of a paper published in year $x$. 
The solid line indicates yearly averages. 
In all figures, $x$ is an ordinary axis and the $y$ axis is logarithmic. 
Figure (a) shows the yearly distribution of raw citation scores 1974 - 2004. 
Figure (b) is the yearly distribution of citation scores normalized by time factor $\tau$, that is $k_t = k / \tau^\beta$. 
Figure (c) is the yearly distribution normalized by time factor $\tau$ and $\phi$ variables, that is $k_{tf} = k / (\tau^\beta \prod_n \phi_n^{\gamma_n})$. 
}
\label{fig:kauthors}
\end{figure}

Figure~\ref{fig:kauthors} plots paper citation score over the number of authors. It is interesting that several papers with seven authors have high citation scores (see the peaks in Figures~\ref{fig:kauthors} a, b, and c). We argue that these are peculiar to the data and may not be generalizable. If we ignore these seven-author papers as exceptions (outliers), there is a general decreasing trend of average citation scores $k$ with an increased number of co-authors.

With normalized citation scores $k_t$ and $k_{tf}$, shown in Figure~\ref{fig:kauthors} (b) and (c), the impact of co-author team size becomes ambiguous. While it is difficult to reach any conclusion about large collaboration teams given the lack of data, many high-impact papers were results of one to three co-authors (see individual data points at the top of Figure~\ref{fig:kauthors} plots). Previous studies have offered evidence about productive collaboration in small teams in InfoVis \citep{ke:infovis,borner:globalbrain}. For example, strong collaboration of three key researchers J.D. Mackinlay, S.K. Card, and G. Robertson, among top scholars listed in Table~\ref{tab:topscholars}, produced several milestone works on information visualization and has been highly regarded in the field.

\section{Conclusion}

We propose a model to analyze citation growth and influences of {\it fitness} variables. Taking into account time factor $\tau$ and prior fitness factors $\phi$, the model offers not only a new formula to predict growing citations but also an approach to quantifying influences of these factors (bias) in scholarly impact analysis. 

Applying the proposed method to modeling paper and scholar citations in the InfoVis 2004 data, a benchmark collection documenting the birth and 31 years' history of information visualization, leads to findings consistent with citation growth in general and our observation about the domain in particular. While $\tau$ and $\phi$ variables have been found to have significant influences on paper citation scores, the overall effect size is considerably large, with $R^2 \approx 0.5$ for the paper fitness model and $R^2 > 0.6$ for the derived scholar fitness model. Citation growth over time follows a power function close to that identified in the scale-free model, in which citation score $k \propto \tau^\beta$ with $\beta=1/2$ \citep{science:rdnnet}. 

Distribution analysis and normalization of citation frequencies based on model estimates provide insights consistent with observations about the domain. Both paper and scholar fitness models reproduce citation frequency distributions that roughly match observed distributions. Isolating the impact of time $\tau$ from raw citation scores produces normalized scores better correlated with a long-term citation benchmark. While plotting raw citation scores over 30 years of InfoVis seems to suggest a counter-intuitive story about the field, normalizing the scores by influences of time and prior fitness reveals a trend consistent with our general understanding of the field. 

Overall, the analysis demonstrates the ability of the proposed model to produce results consistent with the data and to support meaningful comparison of citation scores. The model is based on the general reasoning behind {\it preferential attachment} and {\it fitness} in evolving, growing networks. The simplicity of the proposed fitness modeling, which relies on nothing more than citation records, enables straightforward replication of the reported analysis. We plan to apply the model to analyzing other scientific domains in future studies.


\bibliographystyle{spbasic}  
\bibliography{network}    

\end{document}